# Critical Currents and Melting Temperature of a Two-dimensional Vortex Lattice with Periodic Pinning

L. G. Verga, M. M. Bonilha, M. Carlone, P. A. Venegas

*Departamento de Física, Faculdade de Ciências, UNESP-Universidade Estadual Paulista, CP 473, 17033-360 Bauru-SP, Brazil*

**Abstract** –The critical current and melting temperature of a vortex system are analyzed. Calculations are made for a two dimensional film at finite temperature with two kinds of periodic pinning: hexagonal and Kagomé. A transport current parallel and perpendicular to the main axis of the pinning arrays is applied and molecular dynamics simulations are used to calculate the vortex velocities to obtain the critical currents. The structure factor and displacements of vortices at zero transport current are used to obtain the melting temperature for both pinning arrays. The critical currents are higher for the hexagonal pinning lattice and anisotropic for both pinning arrays. This anisotropy is stronger with temperature for the hexagonal array. For the Kagomé pinning lattice, our analysis shows a multi stage phase melting; that is, as we increase the temperature, each different dynamic phase melts before reaching the melting temperature. Both the melting temperature and critical currents are larger for the hexagonal lattice, indicating the role for the interstitial vortices in decreasing the pinning strength.

## I. Introduction

Many efforts have been made to understand pinning mechanisms in order to obtain higher critical currents. Superconducting films with artificial pinning arrays of well-defined geometry (i.e., square, triangular, rectangular, honeycomb, and Kagomé [1]) have attracted much attention in recent years. Advancements in nanolithographic techniques have permitted the creation of periodic arrays of pinnings and have allowed careful control of the various parameters associated with each system, such as size, shape, and composition. Superconductors with periodic pinning arrays may show vortices that form regular geometrical patterns, depending on the applied external magnetic field (B).

The external magnetic field necessary for the number of vortices to equal the number of pinning sites is called the first matching field, $B_\Phi$. When the vortex lattice is commensurate with the pinning array; that is, when the rate $B/B_\Phi$ is an integer or a rational fraction, the critical current $J_c$ is strongly enhanced. These commensurability effects have been intensively studied [2-9], and they have been observed to reduce the vortex mobility and to help in increasing the critical current. In contrast, for samples where the external magnetic field is increased above the first matching field, some vortices will not be trapped at a pinning site. These interstitial vortices are only weakly pinned in the interstitial regions, and the depinning force decreases faster than it does at commensurate fields. In the present work, we used molecular dynamic simulations to analyze the effects of the Kagomé pinning array ($B/B_\Phi = 4/3$) on the critical current of a two dimensional superconducting film as a function of temperature ($T$), and to compare these effects with those of a hexagonal pinning array ($B/B_\Phi = 1$). We made also a detailed analysis of the structure factor that gives the vortex lattice melting temperature for both pinning arrays. We observe a multi stage melting of the dynamic phases for the Kagomé lattice; that is, each dynamical phase melts at a different temperature, before the melting of the static vortex lattice. Our results show anisotropy in the critical currents and in the dynamical behavior of the vortex system for both pinning arrays, in agreement with experimental results [10,11].

## II. Model

The dynamical properties of the two-dimensional vortex system, under the influence of a periodic pinning network and at finite temperature, can be modeled using a set of Langevin equations. The motion of a vortex at the position **r$_i$** is written as:

$$\eta \frac{d\mathbf{r}_i}{dt} = -\sum_{j \neq i} \nabla_i U_{vv}(r_{ij}) - \sum_p \nabla_i U_p(r_{ip}) + \mathbf{F}_c + f_i^T(t) \quad (1)$$

The first term on the right side of eq. 1 is the vortex-vortex interaction $U_{vv}(r_{ij})=C_v ln(r_{ij}/\Lambda)$, where $r_{ij}$ is the distance between vortex $i$ and $j$ and $C_v$ is the interaction strength. The second term is the vortex pinning interaction $U_p(r_{ip})=-C_p exp(-r_{ip}^2)$, where $r_{ip}$ is the distance between vortex $i$ and pinning $p$, and where $C_p$ is the strength of the vortex-pinning interaction. The third term is the driven force due to transport current **J**, $\mathbf{F}=(\Phi_0/c)\mathbf{J}\times\mathbf{z}$. The last term is the Brownian force due to Gaussian thermal noise, with the properties $<f_i^T(t)>=0$ and $<f_i^T(t)f_j^T(t)>=2\eta k_B T\delta_{ij}\delta(t-t')$. In the present case, we used $\eta=k_B=1$, the length scales normalized by $4\xi$, the time scale normalized by $\tau=16\eta\xi^2/C_v$ and the energy scales by $C_v$. Here, $\xi$ is the coherence length, $\eta$ is the Bardeen-Stephen friction, $\Phi_0$ is the flux quantum, and $\Lambda$ is the effective penetration depth. We perform our calculations for a thin film that is infinite in the transverse ($x$) and longitudinal ($y$) directions and for Kagomé and hexagonal pinning arrays. The critical forces and vortex dynamics are calculated for transport forces in two mutually perpendicular directions, $x$ and $y$, as shown in Figure 1. The numerical simulation is made for the hexagonal pinning array, where the number of vortices, $N_v$, is equal to the number of pinning sites, $N_p$, and for the Kagomé array, where $N_v=4/3N_p$; that is, when the applied magnetic field is equal to the first matching field and 4/3 of the first matching field, respectively.

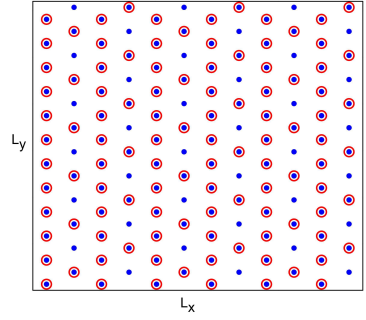

Figure 1. The simulation box, where the dots represent the hexagonal lattice of vortices or the hexagonal array of pinning sites, and the open circles are the Kagomé pinning array.

We analyze the two dimensional system of vortices with periodic pinning at finite temperature by solving the Langevin equations using molecular dynamics simulations. For this purpose, we use a box of size $L_x$ by $L_y$, and periodic boundary conditions to simulate the infinite size of the film. The calculations are made for 144 vortices and a vortex density $n_v=N_v/L_xL_y= 0.12$. For a given temperature, we start the simulation process by relaxing the vortex lattice without the transport current. After 2000 time steps, the most stable configuration is used as the initial boundary condition. The force associated with the transport current is then slowly increased, in steps of $\Delta f=0.01$, up to values as high as $f=4.0$. At each value of force, we take 2000 normalized time steps for equilibration and 30000 time steps for evaluation of the time averages. The temperature is then increased in steps of $\Delta T=0.01$.

The calculation of the critical forces and melting temperature ($T_m$) and the characterization of the different dynamical phases require the calculation of the vortex trajectories, the time averages of vortex velocity, and the structure factor. The critical depinning forces in the $x$ and $y$ directions are determined using the criterion $<V_{x,y}>= 0.001$.

## III. Results and Discussion

In this section, we show the results of velocity, dynamical phases, melting temperature, and critical current for the vortex system interacting with a Kagomé and hexagonal pinning lattice and transport force in *x* and *y* direction.

The analysis of velocities, shown in Figures 2 and 3, together with our results for trajectories and structure factor, clearly demonstrate the different dynamic phases of the vortex system at low temperatures with both pinning arrays. The different strengths of the pinning interaction in the different vortex channels result in a vortex system that shows an anisotropic response when the applied transport force is in the *x* or *y* direction. The analysis of phases for the hexagonal pinning array shows that when the force is applied in the *x* direction, we have two different phases, the static one, which is below the critical current, and the moving phase, which corresponds to a perfect moving crystal. In this case, all vortices start to flow at the same time and move by rectilinear channels along the pinning rows in the *x* direction. When the transport current is in *y* direction there are two different dynamical phases. For values of force slightly higher than the critical current, all vortices start to move at the same time by complex trajectories with short range hexagonal ordering. For higher values of force, the vortices move by winding channels in a more ordered phase.

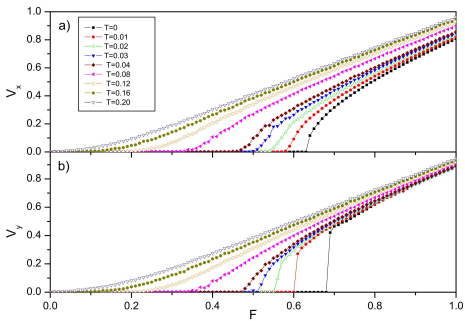

Figure 2: Velocities for the case of a hexagonal pinning lattice at various temperatures as a function of the transport force *F* in: a) the *x* direction and b) the *y* direction.

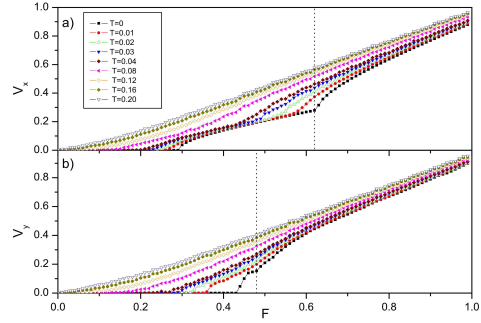

Figure 3: The thermal behavior of velocities for the case of a Kagomé pinning lattice as a function of the driven force *F* in two directions: a) the driven force applied in the *x* direction and b) the driven force applied in the *y* direction. The dotted lines represent the force for the particular case of zero temperature, at which all vortices begin to move. In case a), below the dotted line, half of the vortices are trapped at a pinning site; in case b), a quarter of the vortices are trapped. Above the dotted lines, all vortices move in two sub lattices at different velocities.

For the case of the Kagomé lattice, the behavior is also anisotropic. When the force is applied in the *x* or *y* direction, we have two or three different dynamical phases, respectively. When the force is applied in the *x* direction and the values overcome the critical force, half of the vortices start to move in rectilinear trajectories along the direction of the force and the other half remain trapped. A further increase in the force leads to the second dynamic phase, where all vortices move in rectilinear trajectories but in two sub lattices with different velocities. When the transport force is in the *y* direction and above its critical value, the vortex movement becomes complex, so that in the first dynamical phase, a quarter of the vortices remain trapped and the rest of the vortices move by sinuous channels. Increasing the force causes all of the vortices to move by channels that have interconnectivity resembling the smectic phase found in systems of vortices with random pinning [10]. A further increase in the force causes the vortices to move in well-defined channels over the commensurate rows. This anisotropy in the dynamical behavior of the vortex system is a consequence of the different pinning strengths in the *x* and *y* directions.

As we pointed out before, the Kagomé pinning array results in two sub lattices of vortices that move at different velocities. In Figures 3a and b, we used a vertical dotted line to separate the phase where some of the vortices are pinned and the phase where all vortices are moving, for the particular case of zero temperature. This line is located at $F=0.6$ if the force is applied in the $x$ direction, and $F=0.4$ if the force is applied in the $y$ direction. When the vortex sub lattice containing the remaining trapped vortices starts to move, an abrupt change in the vortex velocity is observed at those values of force. If we look for the thermal behavior of velocity at higher temperatures, we see that this jump in the vortex velocity moves to lower values of force as the temperature is increased, disappearing before it reaches the melting temperature. The thermal fluctuations tend to suppress the pinning effect and when the thermal fluctuations are sufficiently large, the vortex system goes directly from the pinned phase to a phase where all vortices are moving. A multistage melting occurs; that is, prior to the melting of the vortex lattice at zero driven force, there is a melting of the individual dynamical phases at temperatures lower than the melting temperature.

Interestingly, the melting temperature for the static vortex lattice is different for both pinning arrays. We obtain the melting temperature by using the results of the time average power of secondary peaks of the structure factor $S(k)$, and the time average of displacements of vortices $D$, from their equilibrium position at $T=0$. The time average of the structure factor can be written as:

$$S(\boldsymbol{k}) = \langle \left| \frac{1}{N_v} \sum_i \exp\left[i\boldsymbol{k}\cdot\boldsymbol{r_i}(t)\right]\right|^2 \rangle \quad (2)$$

and the time average of the vortex displacements, $D$, can be written as:

$$D = \langle \frac{1}{N_v} \sum_{i=1}^{N_v} |\boldsymbol{r_i}(T) - \boldsymbol{r_i}(0)|^2 \rangle \quad (3)$$

Figures 4 and 5 show the structure factor and displacement as a function of T. Using the criterion $D=L/2$ ($L = \sqrt{L_x^2 + L_y^2}$) for the lattice melting [11], we obtained the melting temperatures $T_m=0.25$ for the hexagonal lattice and $T_m=0.27$ for the Kagomé lattice. These results show clearly that the interstitial vortices are important not only in the dynamical properties of the system but also in the melting of the static vortex lattice. The difference in the melting temperature for each pinning array can be understood because, in the case of the Kagomé array, a quarter of the vortices are interstitial vortices, not trapped at a pinning site. In this case, the thermal fluctuations provoke larger vortex displacements, and hence, a lower melting temperature.

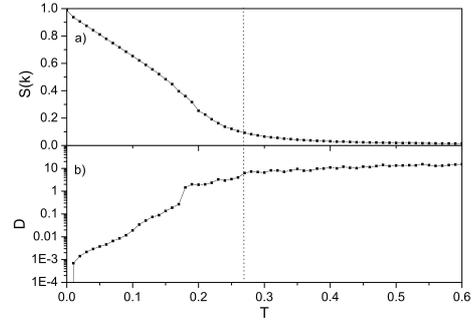

Figure 4: a) The time average power of secondary peaks of the structure factor $S(k)$ and b) the displacement D, for the hexagonal pinning lattice. The dotted line represents the melting temperature.

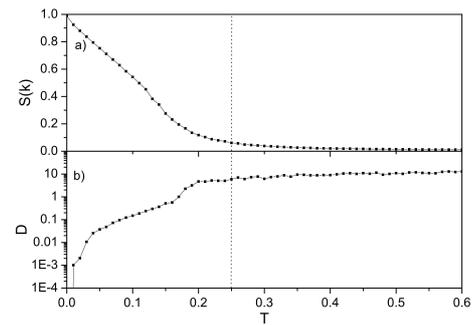

Figure 5: a) The time average power of secondary peaks of the structure factor $S(k)$ of the secondary peaks and b) the displacement $D$, for the Kagomé pinning lattice. The dotted line represents the melting temperature.

From Figures 2 and 3, the critical forces for both pinning arrays and directions of the transport force can be determined with good precision. Figures 6 and 7 show the thermal behavior of the critical forces for the hexagonal and Kagomé lattices, respectively, and the two mutually perpendicular directions of the transport force. From these Figures, we can observe that the critical currents, for all directions of driven force and values of T, are higher for the hexagonal lattice. Figure 1 shows, for the hexagonal lattice, that we have $B/B_\Phi =1$, and all vortices are trapped by a pinning site. In the case of the Kagomé lattice, where $B/B_\Phi=4/3$, pinning vacancies exist and a quarter of the vortices (the interstitial vortices) are not trapped at a pinning site; that is, the lower commensurability leads to a weaker pinning effect and to a lower critical current. Otherwise, for both pinning arrays, the critical forces are anisotropic, and for both pinning arrays, the critical forces in the *y* direction are larger than in the *x* direction.

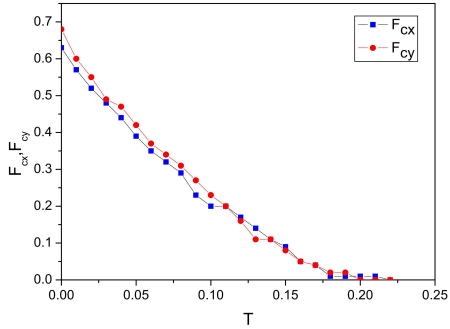

Figure 6: Critical forces ($F_{cx}$, $F_{cy}$) as function of temperature (*T*) for the hexagonal pinning lattice. The driven force in the *x* direction is indicated by squares and in the *y* direction by dots.

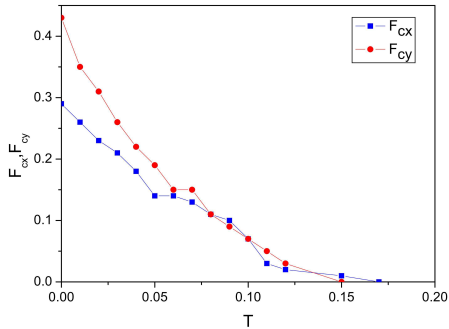

Figure 7: Critical force ($F_{cx}$, $F_{cy}$) as function of temperature (*T*) for the Kagomé pinning lattice. The driven force in the *x* direction is indicated by squares and in the *y* direction by dots.

This anisotropy can be explained by looking at Figure 1. When the vortices are moving in the *x* or *y* directions, they occur as channels with different pinning strengths. Even though the anisotropy is higher for the Kagomé lattice at low temperatures, the anisotropy is stronger with temperature for the hexagonal pinning lattice. This may be understood based on the existence of interstitial vortices. The interstitial vortices in the Kagomé array also mean a lower commensurability and a weaker pinning effect. The thermal fluctuations tend to suppress the effect of pinning, so that systems with weaker vortex-pinning interactions will be more sensitive to these fluctuations.

## IV. Conclusion

We investigated the critical current and melting temperature of a two dimensional vortex system using molecular dynamics simulations. The calculations were performed for hexagonal and Kagomé periodic pinning arrays, where a transport current was applied in two mutually perpendicular directions at a finite temperature. The critical currents and dynamical phases are anisotropic for both pinning networks, and the anisotropy is stronger with temperature for the hexagonal array. Our characterization of dynamic phases shows a multi stage phase melting for the Kagomé pinning lattice, where two vortex sub lattices move at different velocities; that is, as we increase the temperature, each different dynamic phase melts at some point before reaching the melting temperature. The anisotropy in the critical currents and dynamical phases can be explained as a result of the different pinning strengths in the *x* and *y* directions and is in agreement with previous experimental results [10,11]. The higher critical current and melting temperature for the hexagonal lattice can be explained based on its higher commensurability. In the Kagomé lattice, the interstitial vortices play a crucial role; i.e., because they are not seated at a pinning site, the pinning effect is weaker and leads to a larger displacements. This is the reason why hexagonal pinning shows anisotropy at higher temperatures.

## V. Acnowledgments

L.G.V., M.M.B., and M.C. acknowledge Fapesp-Brazil for financial support. This research was supported by resources supplied by the Center for Scientific Computing (NCC/GridUNESP) of the São Paulo State University (UNESP).